\documentclass[aps,prd,onecolumn,groupedaddress,showpacs,nofootinbib,amssymb]{revtex4-2}
\usepackage[dvips]{graphicx}
\usepackage{amssymb}
\usepackage{amsmath}
\usepackage{hyperref}
\usepackage{graphicx,,color}
\usepackage{amsfonts}
\usepackage{bm}
\usepackage{cancel}
\usepackage{comment}
\usepackage{floatflt}
\usepackage{slashed}
\usepackage{appendix}
\usepackage{array}
\usepackage{tabularx}
\usepackage{tikz}

\newcommand\be{\begin{equation}}
\newcommand\ee{\end{equation}}

\allowdisplaybreaks[4]

\begin{document}

\tolerance=5000

\title{Topological Electroweak Phase Transition}
\author{V.K. Oikonomou,$^{1,2,3}$}\email{voikonomou@gapps.auth.gr;v.k.oikonomou1979@gmail.com}
\affiliation{$^{1)}$Department of Physics, Aristotle University of
Thessaloniki, Thessaloniki 54124, Greece \\ $^{2)}$L.N. Gumilyov
Eurasian National University - Astana, 010008, Kazakhstan\\}

 \tolerance=5000

\begin{abstract}
In this work we shall consider the effects of a non-trivial
topology on the effective potential of the Standard Model.
Specifically we shall assume that the spacetime topology is
$S^1\times R^3$ and we shall calculate the Standard Model
effective potential for such a topological spacetime. As we
demonstrate, for small values of the compact dimension radius, the
electroweak symmetry is unbroken, but for a critical length and
beyond, the electroweak symmetry is broken, since the
configuration space of the Higgs field contains an additional
energetically favorable minimum, compared to the minimum at the
origin. The two minima are separated by a barrier, thus a phase
transition can occur, via quantum tunnelling, which mimics a first
order phase transition. This is a non-thermal phase transition,
similar possibly to quantum Hall topological phase transitions,
hence in the context of this scenario, the electroweak symmetry
breaking does not require a high temperature to occur. The present
scenario does not rely on the occurrence of the inflationary era,
only on the expansion of the Universe, however we briefly discuss
the freezing of the superhorizon terms in $S^1\times R^3$
spacetime, if inflation occurred. We also investigate the large
scale differences of the gravitational potential due to the
non-trivial topology. Finally, we briefly mention the distinct
inequivalent topological field configurations that can exist due
to the non-trivial topology, which are classified by the first
Stieffel class which in the case at hand is $H^{1}(S^{1}{\times
R}^{3},Z_{\widetilde{2}})=Z_2$, so even and odd elements can
exist. We also briefly qualitatively discuss how a topologically
induced electroweak phase transition can yield primordial
gravitational waves.
\end{abstract}


\maketitle

\section{Introduction and Executive Summary}

Nowadays cosmology is not a vague theoretical perception for our
Universe, but it is a precision science, based on the plethora of
observational data that are becoming available at present day.
Many cosmological models and theoretical proposals that can
describe the Universe primordially, can and will be tested in the
near future. Indeed, the stage 4 Cosmic Microwave Background (CMB)
experiments \cite{CMB-S4:2016ple,SimonsObservatory:2019qwx} and
the interferometric future gravitational wave experiments
\cite{Hild:2010id,Baker:2019nia,Smith:2019wny,Crowder:2005nr,Smith:2016jqs,Seto:2001qf,Kawamura:2020pcg,Bull:2018lat,LISACosmologyWorkingGroup:2022jok},
are expected to shed light on fundamental problems in theoretical
cosmology, theoretical astrophysics and even high energy particle
physics, due to the sound absence of any particle discovery in the
LHC since the 2012 discovery of the Higgs, for at least two orders
of magnitude (15$\,$TeV center of mass) beyond the Higgs mass
scale. The chorus of observations from the Universe, started with
the astonishing 2017 LIGO-Virgo observation
\cite{TheLIGOScientific:2017qsa,Monitor:2017mdv,GBM:2017lvd,LIGOScientific:2019vic},
and continued in 2023 by the NANOGrav and the Pulsar Timing Arrays
observations of the stochastic gravitational wave background
\cite{NANOGrav:2023gor}. The GW170817 event changed our
theoretical perception of our Universe, since massive gravity
theories of inflation were excluded or severely constrained
\cite{Ezquiaga:2017ekz,Baker:2017hug,Creminelli:2017sry,Sakstein:2017xjx}.
These benchmark points in the history of physics, seem to have
just started an exciting era for theoretical physics and
cosmology. Many theories will be actually experimentally tested on
firm grounds, and this is an important development in the
post-string theory era of theoretical physics, which cannot be
tested on any ground whatsoever.

The uniformization theorem in two and three dimensions basically
classifies the distinct topologies that Riemann surfaces and
hypersurfaces can have in two and three dimensions. In three
dimensions, the classification of the different topologies that
Riemann hypersurfaces can have was done by Thurston
\cite{thurston} and the theorem was proved by Perelman using Ricci
flows on Riemann hypersurfaces
\cite{Perelman:2006up,Perelman:2006un}.

In cosmology, the uniformization theorem basically states that
given a local geometry, the global topology may be non trivial.
Essentially, the local properties of spacetime are determined by
the local metric, but the global properties of spacetime might be
non-trivial and this is determined by the global topology of
spacetime. Cosmologists were continuously concerned for the
spacetime topology and many articles appeared for decades now,
addressing the topology issue in cosmology, see for example Refs.
\cite{deOliveira-Costa:2003utu,Aslanyan:2013lsa,Aslanyan:2011zp,COMPACT:2022nsu,COMPACT:2022gbl,Linde:2004nz,Starobinsky:1993yx,Stevens:1993zz,Vaudrevange:2012da,Goncharov:1987tz,Goncharov:1987qd,Goncharov:1986da,Goncharov:1986tp,Goncharov:1985wb,Aurich:2007yx}
and references therein and also the Planck data
\cite{Planck:2015gmu}.

For quite some time now, the circles-in-the-sky approach in
cosmology had discouraged cosmologists from considering
non-trivial topologies
\cite{Aslanyan:2013lsa,Aslanyan:2011zp,COMPACT:2022gbl}. In fact
if non-trivial topologies existed in spacetime, then the Universe
should repeat itself and thus identical figures of the same
galaxies would appear in the Universe, if the scale of the
non-trivial topology is smaller than the last scattering surface
radius, defined as the distance that light from the decoupling era
travelled up to present day.

Such figures, which would indicate the self-tilling of the last
scattering surface, were never found, thus the topologically
non-trivial Universe was basically abandoned for some years, see
for example Refs.
\cite{Aslanyan:2013lsa,Aslanyan:2011zp,COMPACT:2022gbl} for
discussions on such issues. Recently however it was shown that a
non-trivial topology in the Universe is statistically favorable
compared to a flat Euclidean topology
\cite{Aslanyan:2013lsa,Aslanyan:2011zp,COMPACT:2022gbl}, and also
the anomalies of the Cosmic Microwave Background (CMB) are refined
if a non-trivial spacetime topology exists at scales comparable to
the last scattering surface. Specifically, the toroidal Universes
fit the Planck data in a better way compared to the best-fit
$\Lambda$-Cold-Dark-Matter ($\Lambda$CDM) model, since toroidal
Universes lead to the suppression of the large angular scale CMB
anomalies. This is natural to think since if a non-trivial
topology exists, then the radius of the Universe is finite and not
infinite, thus there is a natural cutoff for the allowed wavemodes
in the CMB. Basically, the largest wavelengths allowed in the CMB
must be smaller that the radius of the non-trivial topological
structure $L$, that is $\lambda \leq L$, or in terms of the
wavenumber, $k\geq \frac{2\pi}{L}$, thus the CMB does not contain
small magnitude modes, or large (infinite) wavelengths. It is
known that small temperature variations in the CMB, including the
anomalous statistical properties of low-multipole harmonic
coefficients, the absence of large-scale correlations and power
asymmetry in the sky, indicate the presence of statistically
anisotropic correlations which can be due to a non-trivial
topology. The presence of a non-trivial topology at scales of the
last scattering surface $\sim H_0^{-1}$, may induce anisotropic
correlations in the CMB temperature and polarization fluctuations.
In fact, this very own feature of the CMB anomalies is one of the
main motivations for having non-trivial topologies in spacetime.
The statistical accuracy of a cosmological model is measured by
the integrated weighted  temperature correlation difference,
$$I=\int_{-1}^{1}\mathrm{d}(\cos \theta)\frac{C^{model}(\theta)-C^{obs}(\theta)}{\mathrm{Var}C^{model}(\theta)}\, ,$$
where $C^{model}(\theta)$ is the two-point temperature correlation
function $C(\theta)=\langle\delta T(\hat{n})\delta
T(\hat{n}')\rangle$ and $\hat{n}\dot \hat{n}'=\cos \theta$. The
values of the integrated weighted  temperature correlation
difference $I$ for toroidal Universes are lower compared to the
best-fit $\Lambda$CDM Universe \cite{Aurich:2007yx} and thus the
toroidal Universes are in better agreement with the observational
data compared to the $\Lambda$CDM Universe. Other approaches, use
the Kullback-Leibler divergence in order to study the statistical
fit of non-trivial topology models \cite{COMPACT:2022gbl}. The CMB
anomalies seem to indicate a preferable length scale in the
Universe which is comparable to $H_0^{-1}$, which could be the
scale of non-trivial topology in the Universe
\cite{COMPACT:2022nsu}. However, it is found that length scales
shorter than the last scattering surface can be observationally
acceptable to a factor of two or even six times shorter
\cite{COMPACT:2022nsu}. Specifically, the Planck 2015 data
indicate that the scale of non-trivial topology must be larger
than $L>6\,H_0^{-1}$, or the radius of the largest inscribed
sphere in the fundamental domain must be $R_i>0.97\,\chi_{rec}$
for the $T^3$ torus and $L>3.5\,H_0^{-1}$ or
$R_i>0.56\,\chi_{rec}$ for $S^1\times R^2$, with the last
scattering surface being $\chi_{rec}=3.1\,H_0^{-1}$ which is of
the order of $14.4\,$Gpc. Theoretical works for the topology
$S^1\times R^2$ indicate that $1.2\leq \frac{L}{\chi_{rec}}\leq
2.1$ \cite{Aslanyan:2011zp}. Later works indicate that at 95.5$\%$
confidence level, the lower bound on the scale of a non-trivial
Universe's topology is $L=1.5 \chi_{rec}$ for a $T^3$ topology,
while for $S^1\times R^2$ we have $L=1.1 \chi_{rec}$
\cite{Aslanyan:2013lsa}.

Now regarding non-trivial topologies, there exist vast classes of
topological structures that can be examined, with curved or flat
three dimensional spacelike hypersurfaces, but for the purposes of
this paper, we limit ourselves to flat topological spaces with
toroidal or circular topology. In fact a two-torus topology
$T^2\times R$ for the spacelike three dimensional hypersurface is
statistically preferable regarding the Planck data
\cite{Planck:2015gmu}, but for simplicity we shall consider only
the topological structure $S^1\times R^2$, the generalization can
easily be done in the topology $T^2\times R$. Compact toroidal
Universes are theoretically motivated because the initial
conditions of low-scale inflationary theories can be explained in
a refined way \cite{Linde:2004nz}. Also the quantum creation of
such compact flat Universes are not exponentially suppressed
compared to curved closed or infinite Universes
\cite{Linde:2004nz}. Primordially, it is expected that before our
Universe became classical, geometry and topology fluctuations have
certainly occurred. In fact topology fluctuations is an essential
and salient feature of quantum gravity theories. Thus once the
Universe became classical, its topology was selected and this
topology might be quite different from a flat Euclidean one.

Regarding the cosmological perspective, apart from the apparent
refinement of the low-multipole anomalies, we shall explore
another important issue in which topology might play an important
role. It is related to the electroweak phase transition, which
basically theoretically relies on the thermal phase transition in
the Higgs sector. In fact the electroweak phase transition is
believed to have occurred during the reheating era, and basically
when the temperature of the Universe reached 100$\,$GeV.
Essentially this is a double phase transition, since from a
horizon temperature perspective, the Universe reached the
100$\,$GeV temperature primordially, with the temperature being
the Gibbons-Hawking temperature $T=H/2\pi$, before the
inflationary era, but the effects of such a phase transition were
diluted by the inflationary expansion. The effects of the
electroweak phase transition during the reheating era can have
observable effects on the stochastic gravitational wave background
spectrum since a stochastic gravitational wave signal is generated
during the electroweak phase transition, peaking at the
frequencies probed by LISA, the Einstein Telescope, DECIGO and the
other future gravitational waves experiments. However, it is
highly questionable whether the Universe has ever reached the
temperature required for the electroweak phase transition to
occur. In fact, there are claims that the so-called reheating
temperature was of MeV order \cite{Hasegawa:2019jsa}. In addition,
the data coming from the NANOGrav and the Pulsar Timing Array
experiments, indicate that an inflationary era with a very low
reheating temperature can explain the NANOGrav signal
\cite{sunnynew}. If this is true in nature, then one must find a
way to explain how the electroweak phase transition have occurred
in the first place. The purpose of this work is to provide a
mechanism for generating the electroweak phase transition non
thermally but via a phase transition due non-trivial spacetime
topology. This mechanism is based on the phase transition in the
Higgs sector due to the non-trivial spacetime topology, for a
spacetime with topology $S^1\times R^3$. Note that we will not
take into account any finite temperature effects, since we assume
that the Universe never reached a reheating temperature large
enough to thermalize the electroweak particles via their
interactions. The phase transition due to the non-trivial
spacetime topology occurs when the radius of the compact dimension
reaches a critical magnitude, after which the electroweak symmetry
is broken. We shall provide a detailed calculation for this phase
transition due to the non-trivial spacetime topology, by
calculating the effective potential of the Higgs sector (Casimir
energy), including its interactions with the gauge bosons, and the
most massive fermions. For the analysis we shall not assume that
inflation occurred, since this scenario is independent of the fact
whether inflation occurred or not, but if inflation occurred, the
phase transition due to the non-trivial spacetime topology
occurred during the inflationary era and near its end possibly,
based on the magnitudes of the critical size of the Universe when
the phase transition occurred. Phase transitions during the
inflationary era have been considered in various different from
ours frameworks, see for example
\cite{Barriga:2000nk,Barriga:2000ma,Adams:1997de,An:2023jxf,Niarchou:2003hz,Labrana:2013oca,Jiang:2015qor},
but most of the effects of such phase transitions are questionable
whether these are diluted eventually by the inflationary
expansion.

This paper is organized as follows: In section II we consider the
effective potential of the Standard Model particles in a spacetime
with topology $S^1\times R^3$ and we present in detail how the
electroweak symmetry breaking takes place as the size of the
compact dimension increases. We also give some hints for the size
of the compact dimension when the electroweak phase transition
took place, which is of the order $L\sim 10^{-21}\,$Km. We also
check the limitations of the perturbative small length expansion
for the Standard Model effective potential in $S^1\times R^3$
spacetime. In section III we validate the fact that superhorizon
inflationary modes in the $S^1\times R^3$ spacetime freeze after
the first horizon crossing, if we assume an inflationary era
during the time that the electroweak phase transition takes place.
In section IV, we consider the large scale modifications of the
gravitational potential in a spacetime with $S^1\times R^3$
topology and we compare this with the Newtonian potential. In
section V we briefly discuss the various distinct field
configurations in a spacetime with $S^1\times R^3$ topology and
indicate the fact that periodic fermionic and anti-periodic scalar
boundary conditions can exist. Finally the conclusions along with
a discussion follow at the end of the paper.

\section{Standard Model Effective Potential in $S^1\times R^3$ and the Electroweak Phase Transition}

\begin{figure}[h!]
\centering
\includegraphics[width=30pc]{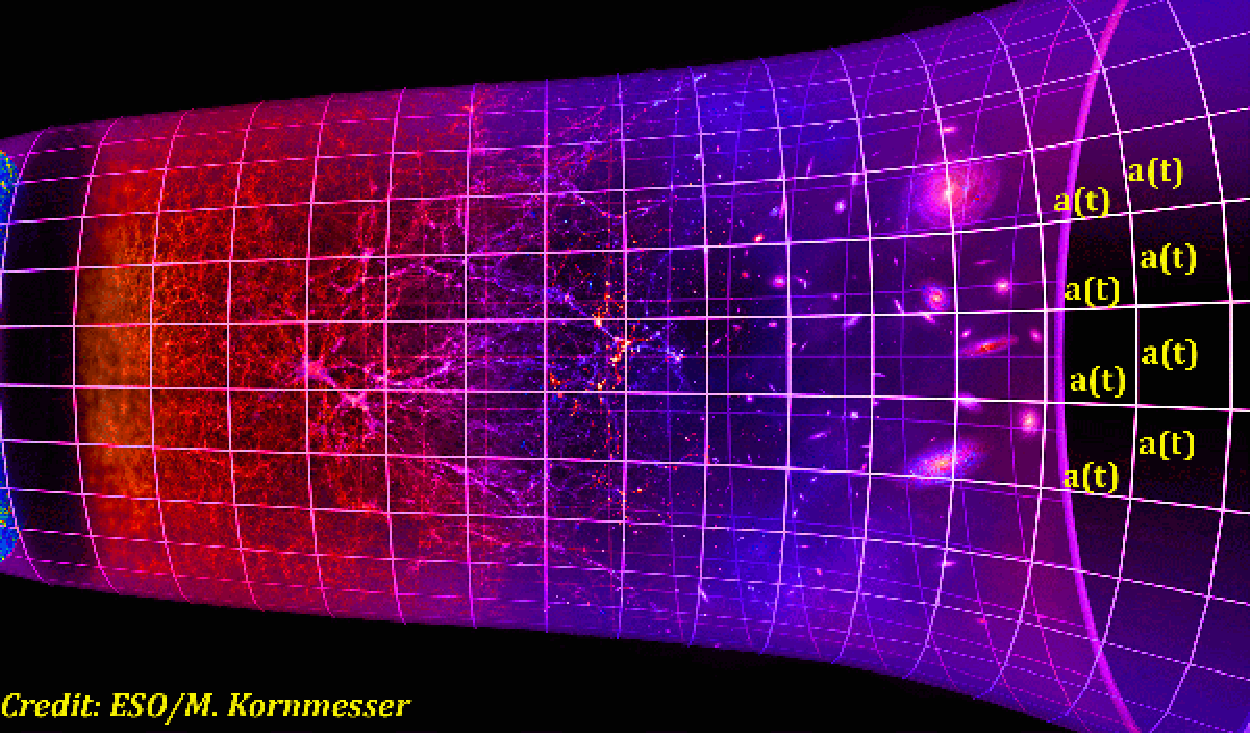}
\caption{Pictorial representation of the spacelike hypersurface of
an $S^1\times R^2$ spacetime. The spacetime locally is described
by a homogeneous and isotropic metric the FRW metric, and the
separations locally in each spatial dimension is given by the
scale factor $a(t)$ which is the same in all spatial dimensions,
while globally one of the two spatial dimensions is a circle. This
figure is edited and it is based on a public image of ESO, which
can be found freely in Credit: ESO/M. Kornmesser:
\url{https://www.eso.org/public/blog/let-there-be-light/}}
\label{plotscale}
\end{figure}

The existence of a compact dimension in the spacetime makes a big
difference compared to the non-compact spacetime case, since the
periodic boundary conditions satisfied by the Fourier perturbation
modes impose an infrared cutoff in the wavenumber and the
wavelengths of the modes. If the radius of the compact dimension
is $L$, wavelengths longer than the radius of the compact
dimensions are not supported thus $k\geq \frac{2\pi}{L}$, which
leads to the suppression of the CMB power at large angular scales
or small multipoles
\cite{Aurich:2007yx,Aslanyan:2013lsa,Aslanyan:2011zp,COMPACT:2022gbl}.
Also the spectrum of the modes is discrete in the case of
non-trivial compact topology. Graphically, the physical
description of the maximum wavenumber allowed in the compact
dimension is depicted below.
\begin{tikzpicture}
  \def\r{1.4cm}
  \def\v{2.5mm}
  \foreach \n in {2,3,4,5,6,12}{
    \begin{scope}[xshift=\n*2*(\r+\v+1mm)]
      \draw[thick]  (0:{\r+\v})
      \foreach \a in {1,...,359}{ -- (\a:{\r+cos(\a*\n)*\v}) } -- cycle;
      \draw[dashed] circle (\r);
    \end{scope}
  }
\end{tikzpicture}
As it can be seen, the maximum wavelength allowed in the compact
spacetime is of the order of the radius of the compact dimension.
Thus primordially, the largest wavelength has the size of the
primordial Universe, thus the length of the compact dimension acts
as a natural cutoff on the wavelengths of the modes allowed. As
the spacetime expands, the maximum wavelength increases, and
nowadays it is of the order of the Hubble horizon of the Universe
$\lambda_{max}\sim H_0^{-1}$. This feature is of great importance,
since wavelengths larger than $\lambda_{max}$ are not allowed in
the spacetime, and equivalently, wavenumbers smaller than
$k_{min}=\frac{2\pi}{\lambda_{max}}$ are not allowed. The maximum
wavelength modes are the ones that primordially had length of the
size of the compact dimension, which eventually expanded and thus
these modes stretched and their wavelength became today comparable
to the Hubble horizon $\sim H_0^{-1}$. This feature is important
since the low-multipole CMB anomalies may be eliminated as we
discussed in the introduction. Once the primordial modes exit the
Hubble horizon during inflation exit the Hubble horizon, thus
becoming superhorizon modes, in principle these should freeze, and
once these re-enter the Hubble horizon during recombination, they
carry unchanged information about the primordial era. This issue
must be examined however, and we will discuss this in a later
section. Let us fix the metric of the spacetime, so we will assume
a flat Friedmann-Lemaitre-Robertson-Walker metric with line
element,
\begin{equation}
\centering \label{metric} ds^2=-dt^2+a^2(t)\delta_{ij}dx^idx^j\, ,
\end{equation}
where $a(t)$ is the scale factor of the Universe. Now an important
issue must be discussed. Although we assumed that the spacetime
has a compact dimension since the topology is non-trivial of the
form $S^1\times R^3$, the spacetime metric has the form of a FRW
spacetime. A natural question to ask whether the non-trivial
topology in one of the three spatial dimensions introduces some
anisotropy. The answer to this question is no and the explanation
is as follows and it is based on a fundamental concept in
Riemannian spaces. The spacetime metric basically describes the
spacetime locally, it is a local property of the spacetime, while
the topology describes spacetime as a whole, that is, it provides
a global description of the spacetime. Thus locally, the spacetime
may be described by the homogeneous metric of Eq. (\ref{metric}),
while globally the spacetime in one or even all the dimensions may
be compact. In fact this is already known from FRW spacetimes with
curvature $K$, in which case the spacetime metric is,
\begin{equation}
\centering \label{metriccurv}
ds^2=-dt^2+a^2(t)\left(\frac{\mathrm{d}r^2}{1-Kr^2}+r^2\mathrm{d}\Omega\right)\,
.
\end{equation}
So in each of the three spatial dimensions, the  scale factor is
$a(t)$ while the spacetime in the case of a positive curvature is
a sphere, thus it is a compact spacetime as a whole. In order to
understand better the local character of the metric and the global
description of the topology, in Fig. \ref{plotscale} we depict the
spatial hypersurface case of $S^1\times R$ spacetime, with a local
FRW metric of the form
\begin{equation}
\centering \label{metric1}
ds^2=-dt^2+a^2(t)\sum_{i=1,2}\delta_{ij}dx^idx^j\, ,
\end{equation}
which is similar to the $S^1\times R^3$ which is hard to depict.
As we can see in Fig. \ref{plotscale}, locally the spacetime is
homogeneous and isotropic, described by the FRW spacetime, while
globally in one dimension the spacetime is compact. Notice that in
all directions, the separations locally in all the two spatial
dimensions are $a(t)$, while globally the spacetime is compact in
one dimension. Now since the spacetime is expanding, the radius of
the compact dimensions which we denote as $L$ is basically time
dependent. Thus $L$, the compact radius is increasing, this is the
same in spirit as in finite-temperature calculations where $T$
decreases as spacetime expands, since the finite temperature
formalism and the compact topology formalism of $S^1\times R^3$
are identical for $L\to 1/T$. Mathematically, the local
description of the spacetime and the global description of the
spacetime is basically known as uniformization theorem and it is
quite important, which was proved for two dimensions, but only
recently was proved for three dimensions by Grisha Perelman using
the Ricci flows formalism \cite{Perelman:2006un}. Our case the
$S^1\times R^3$ spacetime topology is one of the three flat
homogeneous Clifford-Klein spacetimes with topologies
$(S^1)^n\times R^{4-n}$, $n=1,2,3$ which allow a realization of
the FRW flat metric (\ref{metric}), see Refs.
\cite{Goncharov:1985wb,Goncharov:1986da,Goncharov:1986tp,Goncharov:1987qd,Goncharov:1987tz}
for more details and more generalized cases of flat homogeneous
topologically non-trivial spacetimes that admit the FRW metric.
Also the Planck data \cite{Planck:2015gmu} supports our arguments
we discussed here.
Now, if primordially one dimension was compact, then the spectrum
of the primordial scalar and tensor perturbations should have been
discrete, with a lower cutoff wavenumber equal to the inverse of
the length of the compact dimension at that time $k_{min}\sim
\frac{1}{L}$. Thus if an inflationary era occurred, the
perturbations should be treated as not having a continuous
spectrum, but a discrete spectrum, thus the formalism of quantum
primordial perturbation should change. If the perturbations were
generated during the inflationary era, and around the era that the
topological electroweak phase transition occurred, so near a
critical length $L_c$, then the wavenumbers allowed primordially
for the study of the cosmological perturbations are $k_{min}\geq
\frac{2\pi}{L_c}$, so these are too large. The small wavenumbers
of the order of the CMB pivot scale were primordially the modes
having wavelength comparable to the length of the compact
dimension, which stretched and became of the order of the last
scattering surface length when the Universe expanded. It is
exactly the large wavelengths primordially which exited the Hubble
horizon first and became subhorizon during inflation, that are
relevant for CMB physics. In a later section we shall show that
these modes having primordially wavelengths of the same size as
the size of the compact dimension, eventually freeze after the
first horizon crossing and carry information for the primordial
Universe, until they reenter the Hubble horizon during the
recombination era. Thus the study of cosmological perturbations
primordially must include a discrete spectrum of modes with a
minimum cutoff $k_{min}= \frac{2\pi}{L_c}$.

The presence of non-trivial topology induces a vacuum polarization
of all quantum fields \cite{Zeldovich:1984vk}, which is
essentially the gravitational analogue of the Casimir effect. The
corresponding Casimir energy is the vacuum expectation value of
the effective stress energy tensor, or equivalently the effective
potential of the quantum fields with a compact dimension. We shall
consider only one compact dimension and the total spacetime is a
trivial tensor fibre bundle $M=S^1\times R^3$ with the typical
fibers being circles $S^1$ and the base space being $R^3$. Our
approach is to calculate the effective potential of the Standard
Model quantum fields for the spacetime $S^1\times R^3$ imposing
the usual periodic boundary conditions for the bosons and
anti-periodic boundary conditions for the fermions. This is a
rather customary approach inspired by the usual quantum field
theory at finite temperature, but in theories with compact spatial
dimensions, the combination of boundary conditions we used is not
the only one that can be used. We shall come to this issue at a
later section. Given the tree level effective potential of the
theory $V^{tree}(\phi)$ for bosons the effective potential for a
compact circular spatial dimension with radius $L$ is,
\begin{equation}\label{veffbosongeneral}
V_{b}^{eff}=\frac{1}{2L}\sum_{n=-\infty}^{\infty}\int
\mathrm{d}p^3\log\left(p^2+m_{\phi}^2+\left(\frac{2\pi
n}{L}\right)^2 \right)\, ,
\end{equation}
where $m_{\phi}^2=\frac{\partial^2 V^{tree}}{\partial \phi^2}$,
while for fermions with field dependent masses $M_f(\phi)$, the
effective potential is,
\begin{equation}\label{veffbosongeneral}
V_{f}^{eff}=-\frac{1}{L}\sum_{n=-\infty}^{\infty}\mathrm{Tr}\int
\mathrm{d}p^3\log\left(p^2+M_f^2+\left(\frac{(2n+1)\pi}{L}\right)^2
\right)\, .
\end{equation}
The calculation of the above is standard and well known from
finite temperature field theory, by just replacing $T\to 1/L$.
Thus for the Standard Model fields, the one-loop effective
potential at finite volume in the circular dimension, including
the one-loop infinite volume contribution, reads,
\begin{equation}\label{eq:A2}
\begin{split}
    V^{SM}_{eff} (h, L) & = - \frac{\mu^{2}_H}{2} h^2 + \frac{\lambda_H}{4} h^4 + \frac{m^2_h (h)}{24}\frac{1}{L^2} - \frac{1}{12 L \pi} \left[m^2_h (h)\right]^{3/2} + \frac{m^4_h(h)}{64 \pi^2} \left[\ln \left(\frac{a_b}{\mu^2_R\, L}\right) -\frac{3}{2} \right] + \\
    & \frac{3 m^2_{\chi} (h)}{24}\frac{1}{L^2} - \frac{3}{12 \pi\, L} \left[ m^2_{\chi} (h) \right]^{3/2} + \frac{ 3 m^4_{\chi} (h)}{64 \pi^2} \left[\ln \left(\frac{a_b}{\mu^2_R\,L^2}\right) -\frac{3}{2} \right] +  \\
    & \frac{6 m^2_{W} (h)}{24}\frac{1}{L^2} - \frac{4}{12 \pi\, L} m^3_{W} (h) - \frac{2}{12 \pi\, L}\left[ m^2_{W} (h) \right]^{3/2} + \frac{ 6 m^4_{W} (h)}{64 \pi^2} \left[\ln \left(\frac{a_b}{\mu^2_R\,L^2}\right) -\frac{5}{6} \right] + \\
    & \frac{3 m^2_{Z} (h)}{24}\frac{1}{L^2} - \frac{2}{12 \pi\,L} m^3_{Z} (h)  + \frac{ 3 m^4_{Z} (h)}{64 \pi^2} \left[\ln \left(\frac{a_b }{\mu^2_R\,L^2}\right) -\frac{5}{6} \right]  + \\
    & \frac{12 m^2_{t} (h)}{48}\frac{1}{L^2} - \frac{ 12 m^4_{t} (h)}{64 \pi^2} \left[\ln \left(\frac{a_f }{\mu^2_R\,L^2}\right) -\frac{3}{2} \right]
\end{split}
\end{equation}
where \(a_b = \pi^2 \exp{\left(3/2 - 2 \gamma_E \right)}\), \(a_f
= 16 \pi^2 \exp{\left(3/2 - 2 \gamma_E\right)}\), and \(\gamma_E\)
stands for the Euler-Mascheroni constant. The above expression is
a perturbative expansion in terms of $m_i(h)\,L\ll 1$, and this
perturbative expansion must hold true for all the effective masses
that enter the above expression, and it is essentially a small
length expansion which can be the high temperature equivalent
expression. The field dependent masses appearing in Eq.
(\ref{eq:A2}) are,
\begin{equation}\label{eq:A3}
    m^2_h (h) = - \mu^2_H + 3\lambda_H h^2\, ,
\end{equation}
\begin{equation}\label{chimass}
    m^2_{\chi} (h,\phi) = - \mu^2_H + \lambda_H h^2\, ,
\end{equation}
regarding the Higgs and the Goldstone bosons, while for the gauge
fields,
\begin{equation}\label{effectivemassW}
    m^2_W (h) = \frac{g^2}{4} h^2\, ,
\end{equation}
\begin{equation}\label{effectivemassZ}
    m^2_Z (h) = \frac{g^2 + g^{\prime 2}}{4} h^2\, ,
\end{equation}
and for the heaviest quark we have,
\begin{equation}\label{effectivemasstop}
    m^2_t (h) = \frac{y^2_t}{2} h^2
\end{equation}
where \(g\),\(g^{\prime}\) and \(y_t\) are the \(SU(2)_L\),
\(U(1)_Y\) and the top quark Yukawa couplings respectively, and
also the masses are \(m_h = 125\) GeV, \(m_W = 80.4\) GeV, \(m_Z =
91.2\) GeV and \(m_t = 173\) GeV at the current vacuum state of
the Universe (\(h = \upsilon=246\,\)GeV). The focus of this
section is the study of the behavior of the effective potential as
the size of the compact circle dimension varies and to present in
detail the development of the electroweak phase transition as the
length of the compact dimension changes. Let us note that the full
effective potential presented above, can have imaginary
contributions when negative values of the field dependent squared
effective masses occur, as the configuration space variables ($h$)
vary. The terms that may be affected by this issue are the
logarithmic and cubic terms, since the gauge bosons and the
fermions, namely the top quark, always have a  positive squared
effective mass. For the logarithm, the effective mass contribution
is cancelled by the infinite volume one-loop contribution, and we
presented only the resulting expression in Eq. (\ref{eq:A2}). Also
the occurrence of an imaginary part indicates the breakdown of
perturbation theory, see the relevant discussion for high
temperature expansions which are technically equivalent
\cite{Weinberg:1987vp}. Due this this line of reasoning, we shall
consider only the real part of the effective potential, provided
that the imaginary part is insignificant, as in
\cite{Delaunay:2007wb} for high temperature physics.

\begin{figure}
\centering
\includegraphics[width=40pc]{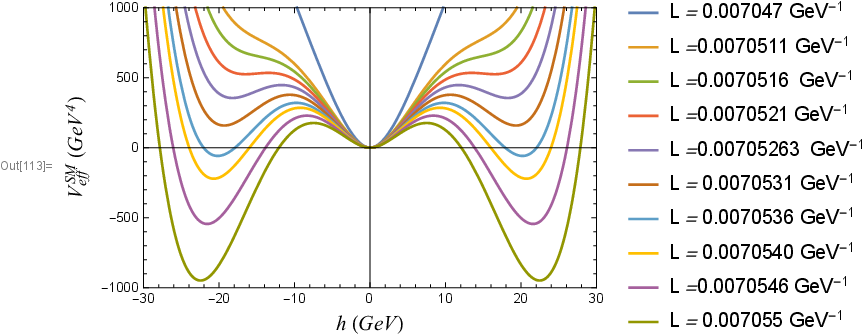}
\caption{The effective potential for the Standard Model particles,
for the spacetime topology $S^1\times R^3$, for various lengths of
the compact circular dimension. As it seems, for small lengths
$L<0.0070511\,$GeV$^{-1}$ or in kilometers $L=1.39139\times
10^{-21}\,$Km, the effective potential is symmetric with only one
minimum at the origin and the electroweak symmetry is unbroken. As
the length increases, a second minimum starts to appear, which
gradually becomes energetically equivalent with the minimum at the
origin, nearly for $L=0.0070536\,$ GeV$^{-1}$ or in kilometers
$L=1.39189 \times 10^{-21}\,$Km. Accordingly the second minimum
becomes energetically favorable for higher values of the magnitude
of the compact dimension and eventually a first order phase
transition can occur because the vacua are separated by a
barrier.}\label{plot1}
\end{figure}

Now let us analyze the topological electroweak phase transition
for the Standard Model effective potential.  In order for the
perturbation expansion to hold true, the condition $m_i(h) L\ll 1$
must be satisfied by all the effective masses defined in Eq.
(\ref{eq:A3})-(\ref{effectivemasstop}), and we shall analyze the
behavior of the perturbative expansion of for various lengths of
the compact dimension and for field values later on in this
section. In Fig. \ref{plot1} we plot the effective potential for
the Standard Model for various lengths of the compact circular
dimension.
As it becomes apparent from Fig. \ref{plot1}, for small lengths
$L<0.0070511\,$GeV$^{-1}$ or in kilometers $L=1.39139\times
10^{-21}\,$Km, the effective potential is fully symmetric, with
only one minimum at the origin existing, and consequently the
electroweak symmetry is unbroken. As the length increases however,
a second minimum starts to appear, which for a critical length of
the compact dimension gradually becomes energetically equivalent
with the minimum at the origin, nearly for $L_c=0.0070536\,$
GeV$^{-1}$ or in kilometers $L=1.39189 \times 10^{-21}\,$Km. As
the Universe expands and the length of the critical dimension
increases, the second minimum becomes energetically more favorable
compared to the local minimum at the origin. The two minima are
separated by a barrier and thus a first order phase transition can
occur via a quantum tunnelling, from the vacuum at the origin to
the energetically favorable electroweak vacuum. The issue of
having a first order phase transition in the Standard Model
effective potential is quite controversial in the literature
concerning finite temperature theories. Indeed initial Standard
Model calculations \cite{Kajantie:1995kf} at finite temperature,
and also recent articles in supersymmetric extensions of the
Standard Model \cite{Wagner:2023vqw,Laine:2012jy}, indicate that
the electroweak phase transition is a second order phase
transition. However, numerous articles using perturbative
calculations indicated that the electroweak phase transition is a
first order phase transition, see for example
\cite{Carrington:1991hz} and references therein. The controversial
character of this issue is due to the mass of the Higgs which is
$\sim 125\,$GeV so the phase transition is even a crossover. The
same arguments should apply in the case of a theory with a compact
dimension by replacing $T\to 1/L$. In this work however, we shall
assume that the phase transition is actually first order, a fact
supported by the barrier separating the inequivalent vacua. Thus
the electroweak symmetry breaking can occur via a phase transition
due to the non-trivial spacetime topology, between the two
inequivalent vacua.

Note that in this scenario in which the electroweak phase
transition occurs due to the non-trivial topology of spacetime and
the change in the radius of the compact dimension, there is no
need for the Universe to reach a high temperature in order for the
electroweak phase transition to occur. Indeed, in standard high
temperature electroweak phase transition scenarios, a high
temperature of $\sim 100\,$GeV is needed in order for the
electroweak phase transition to occur. It is uncertain whether the
Universe reached such a high temperature, thus our scenario
addresses the electroweak phase transition without the need of a
high temperature.

In addition, note that our scenario is not based on the assumption
that an inflationary era ever occurred, the only essential
requirement is the expansion of the Universe. But it is also
possible to incorporate the present scenario into an inflationary
era, and this is rather phenomenologically interesting. Basically,
the size of the compact dimension for which the electroweak
symmetry breaking occurs is in kilometers approximately $L\sim
1.39189 \times 10^{-21}\,$Km, thus the Universe during that era
might be accelerating following an inflationary evolution. Also
note that the scale of the compact dimension when the phase
transition occurs is $L_c=0.0070536\,$ GeV$^{-1}$ but this has
nothing to do with the inflationary scale $H_I\sim 10^{15}\,$GeV,
these are two different phenomena. The length refers to the size
of the compact dimension and this is not related to inflation,
inflation actually affects the rate that the size of the compact
dimension evolves as a function of the cosmic time. Also our
scenario does not specify how the inflationary era might have
occurred, and the inflationary era might be due to some higher
order gravity, or some other simple scalar field scenario, or due
to the existence of a non-trivial vacuum energy due to the compact
dimension. Thus if the electroweak phase transition occurs during
the inflationary era, the induced baryon asymmetry from the first
order phase transition might be diluted due to the accelerating
expansion. However, if the phase transition occurred near the end
of the inflationary era, the induced baryon asymmetry by the phase
transition may not be completely diluted. This is one possibility
which we needed to briefly mention, but we will not further
discuss this issue.

Thus, in this section we showed that is possible to generate the
electroweak phase transition without the need of a high
temperature, just by using a compact topology in one of the three
spatial dimensions. The same calculation of the Standard Model
effective potential can be done for the $T^3$ and $T^2\times R$
topologies, although it would contain more complicated
calculations, but these are well known in the literature as
Selberg zeta function techniques \cite{Elizalde:1997jv}. Of course
one can combine the compact dimension with  finite temperature
corrections, and this scenario is also interesting because a
combined phase transition could be caused by the effects of the
compact dimensions size and the temperature, but we leave these
for future works.

Before we close this section, let us demonstrate the limits of the
validity of the perturbative small length expansion of the
effective potential, which is equivalent to the corresponding high
temperature expansion. The small length expansion requires that
$m_i L\ll 1$ for all the masses participating in the effective
potential. As we showed, the phase transition for the model under
study occurs for lengths of the order $L_c\sim 0.0070536\,$
GeV$^{-1}$ or in kilometers $L=1.39189 \times 10^{-21}\,$Km, so a
question is whether the small length expansion holds true near the
phase transition, and when does the perturbative expansion no
longer holds true.
\begin{figure}
\centering
\includegraphics[width=20pc]{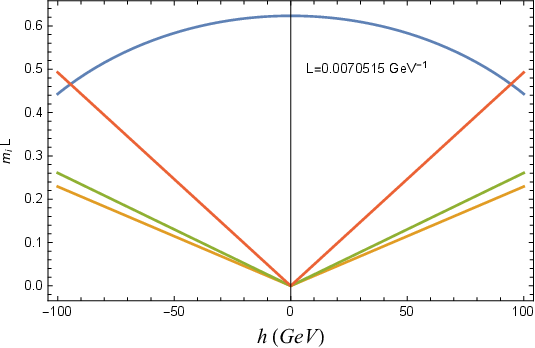}
\includegraphics[width=20pc]{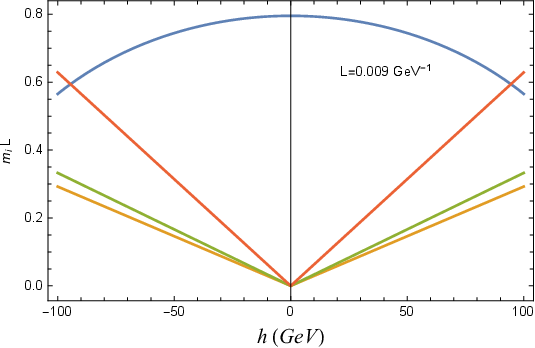}
\includegraphics[width=20pc]{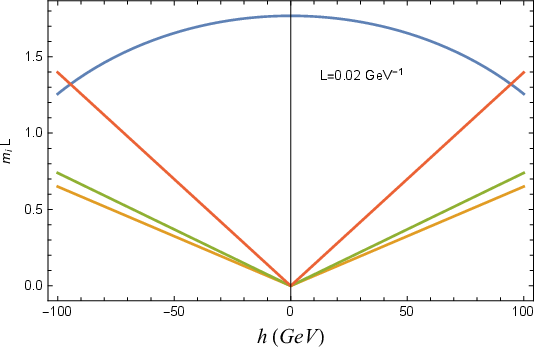}
\caption{Numerical examination of the small length perturbation
expansion for the effective potential. The plots correspond to the
terms $m_i L$ for three typical lengths, $L=0.0070516\,$GeV$^{-1}$
($L=1.39149 \times 10^{-21}\,$Km), $L=0.009\,$GeV$^{-1}$
($L=1.77597 \times 10^{-21}\,$Km) and $L=0.02\,$GeV$^{-1}$
($L=3.9466 \times 10^{-21}\,$Km), with $i=h,W,Z,t$.}\label{plot2}
\end{figure}
In Fig. \ref{plot2} we plot the relevant terms $m_i L$ for three
typical lengths, $L=0.0070516\,$GeV$^{-1}$ ($L=1.39149 \times
10^{-21}\,$Km), $L=0.009\,$GeV$^{-1}$ ($L=1.77597 \times
10^{-21}\,$Km) and $L=0.02\,$GeV$^{-1}$ ($L=3.9466 \times
10^{-21}\,$Km), with $i=h,W,Z,t$. As it can be seen, the small
length perturbation expansion essentially breaks for
$L=0.02\,$GeV$^{-1}$, but note that the phase transition has
occurred and is considered already completed for smaller lengths.
Hence, the small length perturbation expansion describing the
effective potential during the first order phase transition is
valid and hence the calculations are valid.

\section{Superhorizon Inflationary Tensor Modes in the Spacetime with $S^1\times R^3$ Topology}

Let us now assume that the inflationary era took place and all the
spatial dimensions of spacetime inflated, including the compact
one. The question is what would happen with the cosmological
perturbations if a compact dimension exists? Apparently, plainly
spoken, the integrals are turned to sums, or more formally, the
wavenumbers of the cosmological perturbations are not continuous
anymore but discrete. Also, as we mentioned earlier, the fact that
a compact dimension exists indicates that there is a maximum
wavelength allowed in the compact dimension, and thus a minimum
wavenumber. Assuming that the compact dimension has a length of
the order of the critical length for which the electroweak phase
transition occurred, this means that the wavenumber during
inflation must have been larger than $k\geq
\frac{2\pi}{L_c}=\,$Mpc$^{-1}$. Once the inflationary era
commences, the Hubble horizon shrinks and the cosmological
perturbations with wavenumbers $k\geq
\frac{2\pi}{L_c}=\,$Mpc$^{-1}$ exit the Hubble horizon. In this
section we briefly discuss the evolution of these tensor
superhorizon modes for a de Sitter era, for a Universe with
$S^1\times R^3$ topology. In this way we shall have a concrete
idea on the effects of the discretization of the modes occurred by
the compact dimension. We shall consider only the tensor
perturbations. We shall assume an Einstein-Hilbert gravity without
any modified gravity generating inflation, for example a scalar
field generated inflationary era. We consider the following tensor
perturbations of the flat FRW metric,
\begin{equation}\label{tensorpert}
\mathrm{d}s^2=-\mathrm{d}t^2+a(t)^2\left(\delta_{ij}+h_{ij}
\right)\mathrm{d}x^i\mathrm{d}x^j\, ,
\end{equation}
and using the Fourier transform of the tensor perturbation $h_{i
j}$,
\begin{equation}\label{tensorperturbationfourier}
h_{i j}(\vec{x},t)=\sqrt{V}\int
\frac{\mathrm{d}^3k}{(2\pi)^3}\sum_{\ell}\epsilon_{i
j}^{\ell}h_{\ell k}e^{i\vec{k}\vec{x}}\, ,
\end{equation}
with ``$\ell$'' denoting the polarization of the tensor
perturbation, and $V$ denotes the volume element. The Fourier
transformation of the tensor perturbation $h_{i j}$ satisfies,
\begin{equation}\label{mainevolutiondiffeqnfrgravity}
\ddot{h}_{\ell}(k)+\left(3+\alpha_M
\right)H\dot{h}_{\ell}(k)+\frac{k^2}{a^2}h_{\ell}(k)=0\, ,
\end{equation}
with $\alpha_M$ being defined as,
\begin{equation}\label{amfrgravity}
a_M=\frac{\dot{Q}_t}{Q_tH}\, ,
\end{equation}
which is zero for an Einstein-Hilbert gravity. Thus the resulting
Mukhanov-Sasaki equation for the tensor perturbations is,
\begin{equation}\label{mainevolutiondiffeqnfrgravity1}
\ddot{h}_{\ell}(k)+3H\dot{h}_{\ell}(k)+\frac{k^2}{a^2}h_{\ell}(k)=0\,
,
\end{equation}
so if we make the wavenumbers discrete, then the minimum
wavenumber allowed is $k=\frac{2\pi}{L}$. Assume that
$L=L_c=\,$Mpc$^{-1}$, thus we will study the evolution of such
perturbations. When such modes become subhorizon during the de
Sitter era, we have $\frac{2\pi}{L_c}\gg a_I H_I$, where $H_I$ is
the Hubble rate during inflation and $a$ is the scale factor. For
a de Sitter evolution, the solution of Eq.
(\ref{mainevolutiondiffeqnfrgravity1}) with $k= \frac{2\pi n}{L}$
is the following,
\begin{equation}\label{solutionofsuperhorizon}
h_{\ell}(t)=c_1\frac{\pi  n \sqrt{e^{-2 H_I t}} \sin
\left(\frac{\pi n \sqrt{e^{-2 H_I t}}}{H_I L}\right)}{H_I
L}+c_1\cos \left(\frac{\pi n \sqrt{e^{-2 H_I t}}}{H_I L}\right)\,
,
\end{equation}
where $c_1$ is an integration constant. Thus apparently the tensor
perturbations after crossing the horizon decay exponentially and
freeze, even for the smallest wavenumber with $n=1$. Note that all
the polarization modes propagate equivalently in spacetime so they
satisfy the same equation (\ref{mainevolutiondiffeqnfrgravity1}).

\section{Non-trivial Topology Effects of Gravitational Interactions at Large Scales}

Now an important question that is raised is what would be the
gravitational effects of a compact dimension at galactic scales.
In this section we shall discuss this issue in some detail. The
modifications of Newton's gravity due to an $S^1\times R^2$
spatial topology was addressed in Ref. \cite{Floratos:2012nf}.
Following Ref. \cite{Floratos:2012nf}, the gravitational potential
for a spacetime with spatial topology $S^1\times R^2$ has the
following general form (with the Newton's gravitational constant
set to unity),
\begin{equation}\label{gravitationalpotentialgeneral}
\Phi(r,L,\theta)=-\frac{1}{4\pi
\sqrt{r^2+L^2q^2}}-\frac{1}{4\pi}\sum_{\ell=1}^{\infty}\left(\frac{1}{4\pi
\sqrt{r^2+L^2(2\pi \ell-q)^2}}-\frac{1}{2\pi \ell L}+\frac{1}{4\pi
\sqrt{r^2+L^2(2\pi \ell+q)^2}}-\frac{1}{2\pi \ell L} \right)\, ,
\end{equation}
where $q=[0,2\pi)$. The anisotropy introduced by a compact
dimension is quantified by the presence of the parameter $q$. Now
if the distance $r$ from the gravitational center is quite smaller
than the size of the compact dimension $L$, that is for $r/L\ll
1$, the gravitational potential takes the form
\cite{Floratos:2012nf},
\begin{equation}\label{finalextnesion}
\Phi(r,L,\theta)=-\frac{1}{4\pi\,r}\left(1+2\sum_{n=1}^{\infty}\zeta
(2n+1)\left(\frac{r}{2\pi L} \right)^{2n+1} P_{2n}(\cos \theta)
\right)\, ,
\end{equation}
taking into account that $L\,q=r\cos \theta$, which holds true
when $r/(2\pi L)\ll 1$. Now let us consider the gravitational
potential for $\theta=0$, in which case $P_{2n}(0)=1$ and let us
try to present pictorially which is the difference between the
$R^3$ and $S^1\times R^2$ topology in the gravitational potential.
We shall assume that $L=1.1 \chi_{rec}$ where recall that
$\chi_{rec}$ is the last scattering surface distance and also we
shall consider only the first three terms in the sum in Eq.
(\ref{finalextnesion}).
\begin{figure}
\centering
\includegraphics[width=20pc]{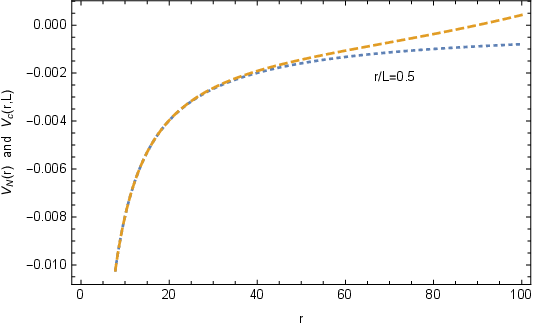}
\includegraphics[width=20pc]{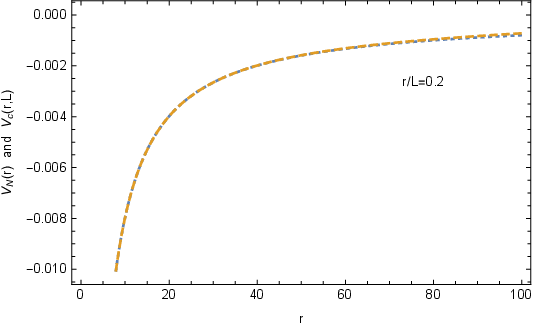}
\includegraphics[width=20pc]{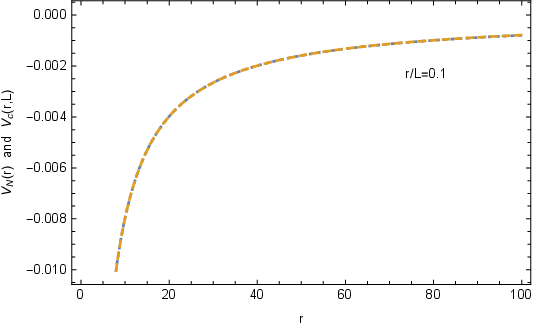}
\caption{Numerical examination of the small length perturbation
expansion for the effective potential. The plots correspond to the
terms $m_i L$ for three typical lengths,
$L=0.0070516\,$GeV$^{-1}$, $L=0.009\,$GeV$^{-1}$ and
$L=0.02\,$GeV$^{-1}$, with $i=h,W,Z,t$.}\label{plot3}
\end{figure}
In Fig. \ref{plot3} we plot the Newtonian potential (dotted curve)
versus the modified gravitational potential (dashed curve) in
arbitrary units, for distances in which the maximum considered
distance $r$ is chosen to satisfy $\mathrm{max} (r/L)=0.5$ and
$L=200$ (so for distances quite close to the last scattering
surface), $\mathrm{max} (r/L)=0.2$ and $L=500$ and finally
$\mathrm{max} r/L=0.1$ and $L=1000$. The only case for which
differences can be found is when the distances considered are
quite close to the last scattering surface, $\mathrm{max}
(r/L)=0.5$ and $L=500$. This is a rather qualitative approach
without true distances being considered, however it is apparent
that if the compact dimension has radius $L=1.1 \chi_{rec}$, then
even for large cosmological distances, the Newtonian potential
perfectly describes the gravitational potential and no observable
effects can be detected. This said behavior is well supported by
observations \cite{White:2001kt}.

\section{Non-trivial Topology Implications for Boundary Conditions and Non-trivial Particle Configurations for the Topology $S^1\times R^3$}

As a final consideration for this paper we will discuss the
possible configurations that are allowed due to the non-trivial
topology of the spacetime. The boundary conditions allowed for
spacetimes with non-trivial topologies are different compared to
topologically trivial spacetimes. Twisted fields are allowed, and
this was first realized by Isham \cite{isham} and then it was
adopted in other works \cite{Goncharov:1986tp,ford,spalluci}. For
the case at hand, the topological properties of the $S^{1}{\times
R}^{3}$ spacetime are determined by the first Stieffel class
$H^{1}(S^{1}{\times R}^{3},Z_{\widetilde{2}})$ and the isomorphism
if the first Stieffel class to the singular (simplicial)
cohomology group ${H}_{1}({S} ^{1}{\times R}^{3}{,Z}_{2})$, due to
the fact that the ${Z}_{\widetilde{2}}$ sheaf is trivial. The
first Stieffel class $H^{1}{(S}^{1}{\times
R}^{3}{,Z}_{\widetilde{2}}{)=Z}_{2}$ determines the twisting of a
fiber bundle and specifically it determines the global
orientability of the fiber bundle. For the case at hand, the
orientability is classified by the  ${Z}_{2}$ group, thus there
are two locally equivalent fiber bundles, which are globally
different, a cylinder-like and a Moebius strip-like bundle. The
sections of these globally distinct fiber bundles are scalar
fields and fermions, which are charged with a topological charge
called Moebiosity, or also known as twist, and is given according
to the first Stieffel class $H^{1}{(S}^{1}{\times
R}^{3}{,Z}_{\widetilde{2}}{)=Z}_{2}$. The twisted fields satisfy
periodic or anti-periodic boundary conditions in the compact
dimension, and there is no discrimination on the boundary
conditions that spinor sections or scalar sections can obey. Thus
we may have periodic boundary conditions for fermions (twisted
fermions) or even anti-periodic boundary conditions for scalars
(twisted scalars). This is in contrast to finite temperature field
theory considerations in which fermions satisfy anti-periodic
boundary conditions and bosons satisfy periodic boundary
conditions. The only difference is that twisted fields cannot have
a vacuum expectation value. The total Lagrangian has to have
Moebiosity zero, thus the interaction vertices must all be
trivial. Then if $\varphi _{u}$, $\varphi_{t}$ stand for the
untwisted and twisted scalar respectively, and $\psi _{t}$, $\psi
_{u}$ stand for twisted and untwisted spinor respectively, then
the allowed boundary conditions in the compact ${S}^{1}$ dimension
are, $\varphi
_{u}(x,0)=\varphi _{u}(x,L)$ and $\varphi _{t}(x,0)=-\varphi _{t}(x,L)%
$ for the scalar fields and $\psi _{u}(x,0)=\psi _{u}(x,L)$, $\psi%
_{t}(x,0)=-\psi _{t}(x,L)$ for the fermionic fields, with $x$
denoting the remaining spacetime dimensions. Let the untwisted
fields twist be denoted by ${h}_{0}$  (which is a trivial element
of the ${Z}_{2}$ group) and also the twist of the twisted fields
is denoted by $h_{1}$ (which is the non
-trivial element of the  ${Z}_{2}$ group). Now $h_{0}+h_{0}=h_{0}$ ($0+0=0$), $%
h_{1}+h_{1}=h_{0}$ ($1+1=0$), $h_{1}+h_{0}=h_{1}$ ($1+0=1$).
Hence, the topological charges at the
interaction vertices yield a total sum ${h}_{0}$ under the ${H}^{1}{(S}^{1}{\times R}^{3}{,Z%
}_{\widetilde{2}}{)}$. In our case, many combinations can be made
assigning a non-trivial twist to some fermions or even to some
gauge bosons, and in principle this could affect the critical
length for which the phase transition occurs and also the overall
vacuum energy, even changing in sign. However the constraint that
the Standard Model Lagrangian must have Moebiosity $h_0$, narrows
down the non-trivial topological fields. Let us briefly discuss
this issue. Regarding the fermions in the Standard Model
Lagrangian, they appear in the kinetic terms,
\begin{equation}\label{kinetiterm}
\mathcal{L}\sim
\sum_i\bar{\psi}_i\gamma^{\mu}\partial_{\mu}\psi_i\, ,
\end{equation}
the Yukawa couplings,
\begin{equation}\label{yukawa}
\mathcal{L}\sim -y_e\left( \bar{\psi}_L\Phi\psi_R+
\bar{\psi}_R\Phi\psi_L\right)\, ,
\end{equation}
and the gauge interactions,
\begin{equation}\label{gaugeinteractions}
\mathcal{L}\sim \sum_i
\bar{\psi}_i\gamma^{\mu}\left(i\partial_{\mu}-\frac{g'}{2}Y_W
B_{\mu} -\frac{g}{2}Y_W W_{\mu}^a\tau_a\right)\psi_i\, ,
\end{equation}
while the gauge bosons also have the kinetic terms,
\begin{equation}\label{gaugebosonsinteractions}
\mathcal{L}\sim -\frac{1}{4}B_{\mu \nu}B^{\mu \nu}
-\frac{1}{2}\mathrm{tr}W_{\mu \nu}W^{\mu
\nu}-\frac{1}{2}\mathrm{tr}G_{\mu \nu}G^{\mu \nu}\, .
\end{equation}
Thus, since the Higgs particle must have trivial Moebiosity, it is
apparent that the gauge bosons must also have trivial Moebiosity
due to their appearance in the gauge interactions with the
fermions. Hence, the only particles that can have non-trivial
Moebiosity are the fermions and possibly the Goldstone bosons. The
full analysis with periodic fermions and anti-periodic Goldstone
bosons will be done in a future work.

\section{Discussion of the Results and Future Perspectives}

In this work we presented a scenario in which the electroweak
phase transition may occur topologically. In this scenario, the
spacetime topology is $S^1\times R^3$ and the electroweak phase
transition is triggered topologically due to the finite volume
effects induced by the compact dimension. We calculated the
effective potential of the Standard Model particles by taking into
account the finite volume effects induced by one compact
dimension. As we demonstrated the electroweak symmetry is unbroken
for small values of the compactification length, however as the
radius of the compact dimension increases, the electroweak
symmetry breaks via a first order phase transition. Indeed, for
small values of the radius of the compact dimension, the effective
potential has only one minimum at the origin, but as the length of
the compact dimension increases, then a second local minimum is
developed, which gradually becomes energetically equivalent to the
minimum at the origin and eventually becomes energetically
favorable after a critical value of the radius of the compact
dimension. The two inequivalent minima are separated by a barrier,
thus it is possible that the Universe may transit from the vacuum
at the origin to the energetically favorable minimum via a first
order phase transition, a quantum tunnelling controlled by finite
volume effects. We also examined the limitations of the small
length expansion in the effective potential, which is equivalent
to the high temperature expansion in thermal field theory.

The attribute of having a topologically induced electroweak phase
transition is that there is no need for a high reheating
temperature. Indeed, it is questionable if the Universe ever
reached such high temperatures needed for the electroweak symmetry
breaking to occur thermally, thus this phase transition due to the
non-trivial spacetime topology evades the standard thermal
electroweak phase transition.

As we demonstrated, it is possible that the Universe might quantum
tunnel from the minimum at the origin to the energetically
favorable minimum, via a topological first order phase transition.
This process is a non-equilibrium process, thus one of the
Sakharov criteria is satisfied and therefore this process can lead
to baryon asymmetry in the Universe. However, due to the fact that
the critical magnitude of the compact dimension for which the
inequivalent minima occur is of the order $\sim
\mathcal{O}(10^{-21})$Km, it is possible that, if inflation
occurred, this whole phase transition due to the non-trivial
spacetime topology took place during the inflationary era. Thus,
any baryon asymmetry induced by the phase transition due to the
non-trivial spacetime topology may have been washed away. We did
not go in much details on this issue though, and this feature is
more or less dependent on whether inflation occurred or not. Now
assuming that inflation occurred, we demonstrated that the
superhorizon modes in the $S^1\times R^3$ spacetime indeed freeze
after the first horizon crossing. Also we investigated the effects
of the compact dimension in the gravitational potential and we
compared the $S^\times R^2$ potential to the $R^3$ Newtonian
potential. Finally, we briefly mentioned the distinct topological
field configurations allowed in spacetimes with a compact space
dimension. As we discussed, it is possible to have anti-periodic
bosons and periodic fermions in the compact dimension, in contrast
to thermal physics in which one may have solely periodic bosons
and anti-periodic fermions.

Some interesting questions arise as a result of this work.
Regardless if inflation occurred or not, such a topologically
induced phase transition may actually be the source of primordial
gravitational waves. This phase transition is not thermal however,
and has to do with a topological change in the configuration field
space of the Higgs particle. A phase transition due to the
non-trivial spacetime topology in the configuration space refers
to a change in the global properties or structure of a physical
system as it undergoes a transition between different phases, in
our case the two inequivalent Higgs vacua that occur after a
critical length of the compact dimension, where these phases are
characterized by distinct topological properties of the
configuration space (broken and unbroken electroweak symmetry).
One example of a phase transition due to the non-trivial spacetime
topology is the transition between different topological phases of
matter, such as the transition between different quantum Hall
states in two-dimensional electron systems or the transition
between different topological insulator phases. These transitions
are characterized by changes in the topology of the band structure
or the emergence/disappearance of topologically protected edge or
surface states. The transition between different quantum Hall
states is not a thermal transition in the conventional sense,
meaning it's not primarily driven by changes in temperature
\cite{cond}. Instead, it's a quantum phase transition, which
occurs at absolute zero temperature ($T = 0\,$K) due to changes in
parameters such as magnetic field strength, electron density, or
disorder in the material. It is questionable if the bubble
nucleation approach can be applied in our case, although in
principle the formalism can be applied
\cite{Kamionkowski:1993fg,Weir:2016tov,An:2022cce,Huber:2008hg}.
Indeed, bubble nucleation can be used to illustrate some aspects
of a phase transition due to the non-trivial spacetime topology,
particularly in the context of phase transitions involving changes
in the topology of the configuration space, in our case the
configuration space of the Higgs field. In the study of phase
transitions, what we call as bubble nucleation usually refers to
the spontaneous formation of localized regions (bubbles) of a new
phase (new minimum in the Higgs field) within an initial parent
phase (the electroweak symmetry minimum at the origin). This phase
transition occurs when the system reaches a critical point,
quantified by a relevant quantity that its change triggers the
change in the configuration space, in our case the critical length
of the compact dimension. One way to understand how bubble
nucleation can relate to a phase transition due to the non-trivial
spacetime topology is through the concept of symmetry breaking. In
many phase transitions, the transition between different phases is
associated with the breaking of certain symmetries of the system.
Topological phase transitions often involve the breaking or
changing of symmetries, leading to the emergence or disappearance
of certain topological properties. For example, consider the
transition between a normal phase and a superconducting phase in a
material. In the normal phase, the system exhibits certain
symmetries, while in the superconducting phase, these symmetries
are broken due to the formation of Cooper pairs and the onset of
superconductivity. This transition can be understood as a phase
transition due to the non-trivial spacetime topology, as it
involves changes in the topology of the system's configuration
space. When a phase transition due to the non-trivial spacetime
topology occurs, localized regions of the new phase can nucleate
within the parent phase, similar to how bubbles form during bubble
nucleation. These nucleated regions grow and eventually percolate
throughout the system, driving the system into the new phase.
While bubble nucleation itself may not fully capture all aspects
of a phase transition due to the non-trivial spacetime topology,
it provides a useful conceptual framework for understanding the
formation and growth of new phases within a system undergoing a
phase transition due to the non-trivial spacetime topology. In
addition, the study of bubble nucleation in the context of phase
transitions has contributed valuable insights into the dynamics
and kinetics of phase transitions, which are relevant to
understanding phase transitions due to the non-trivial spacetime
topology as well. Thus, it would be quite interesting to study the
generation of gravitational waves during such a non-equilibrium
era in our Universe. The gravitational waves should correspond to
a primordial era of our Universe, and thus could be probed by LISA
and other future interferometers. Note that in our approach, there
is no need for an inflationary era, the only necessary ingredient
for the phase transition due to the non-trivial spacetime topology
to work is the expansion of the Universe. Another interesting
perspective is to include high temperature effects, and also to
study other topological configurations of the spacelike
hypersurface of the spacetime, such as $T^3$ and $T^2\times R$
topologies, which are also theoretically and statistically
motivated, see the relevant text in the introduction. The
calculations in this case would be a bit more complicated, since
Selberg zeta function techniques \cite{Elizalde:1997jv} would be
involved, but the problem is feasible and interesting to develop.
We hope to address this issue in a future work. Also invoking a
combination of a topological and a thermal phase transition would
also be interesting for phenomenological reasons, and also
discussing the effects of a phase transition due to the
non-trivial spacetime topology during the inflationary era. We
leave such conceptually interesting tasks for future studies.

\section*{Acknowledgments}

This research has been is funded by the Committee of Science of
the Ministry of Education and Science of the Republic of
Kazakhstan (Grant No. AP19674478) (V.K. Oikonomou).

\end{document}